\documentclass[letterpaper, 10 pt, conference]{ieeeconf}

\IEEEoverridecommandlockouts

\usepackage{amsmath,amsfonts}
\usepackage{graphicx}
\usepackage{textcomp}
\usepackage{comment}

\usepackage{amsthm}
\usepackage[utf8]{inputenc}
\usepackage[T1]{fontenc}
\usepackage{enumerate}
\usepackage{soul}
\usepackage[dvipsnames,table]{xcolor}
\usepackage[hidelinks]{hyperref}
\usepackage{siunitx}
\usepackage{multirow}
\usepackage{subfig}
\usepackage{booktabs}
\usepackage[ruled,vlined]{algorithm2e}
\usepackage{algpseudocode}
\DontPrintSemicolon

\usepackage{amssymb}
\usepackage{mathtools}
\usepackage{bm}
\usepackage{accents}
\usepackage{caption}
\usepackage{colortbl}
\usepackage{array}
\usepackage{adjustbox}
\usepackage{wasysym}
\usepackage{pifont}
\usepackage{stfloats}
\usepackage{url}
\usepackage{verbatim}
\usepackage{makecell}
\usepackage{tikz}
\usepackage{color}
\usepackage{lipsum}
\usepackage{float}

\usepackage[style=ieee,backend=biber,sortcites=true]{biblatex}

\usepackage{fontawesome5}

\renewcommand*{\multicitedelim}{\addcomma\space}
\DeclareCiteCommand{\cite}[\mkbibbrackets]
  {\usebibmacro{prenote}}
  {\usebibmacro{citeindex}\printfield{labelnumber}}
  {\multicitedelim}
  {\usebibmacro{postnote}}
  
\AtBeginBibliography{\scriptsize}
\definecolor{blue}{HTML}{4E95D9}
\definecolor{gold}{HTML}{E5AE1C}
\definecolor{safe}{HTML}{4169E1}
\definecolor{unsafe}{HTML}{C41230}

\newtheorem{proposition}{Proposition}

\providecommand{\R}{\mathbb{R}}

\DeclareMathOperator*{\argmin}{argmin}

\newcommand{\dsmall}{\small\baselineskip=12pt\relax}

\definecolor{myblue}{HTML}{0072B2}
\definecolor{orange}{HTML}{D55E00}

\AtEveryBibitem{%
  \clearfield{doi}\clearfield{url}\clearfield{urlyear}\clearfield{urlmonth}\clearfield{urlday}%
  \clearfield{isbn}\clearfield{issn}\clearfield{eprint}\clearfield{month}%
  \clearlist{language}%
  \ifentrytype{inproceedings}{%
    \clearlist{publisher}\clearlist{organization}\clearlist{location}%
    \clearname{editor}\clearfield{series}\clearfield{volume}}{}%
}

\author{Maitham F. AL-Sunni$^\ast$ \quad Turki Bin Mohaya$^\ast$ \quad John M. Dolan \quad Peter Seiler%
\thanks{$^\ast$ These authors contributed equally to the work.}%
\thanks{M. F. AL-Sunni is with the Department of Electrical \& Computer Engineering, Carnegie Mellon University, Pittsburgh, PA, USA. Email: {\tt malsunni@andrew.cmu.edu}}%
\thanks{T. Bin Mohaya and P. Seiler are with the Department of Electrical Engineering and Computer Science at the University of Michigan, Ann Arbor, MI, USA. Email: {\tt \{turki,pseiler\}@umich.edu}}%
\thanks{J. M. Dolan is with the Robotics Institute, Carnegie Mellon University, Pittsburgh, PA, USA. Email: {\tt jdolan@andrew.cmu.edu}}%
}

\begin{document}

\title{\LARGE \bf Generalizable Output Feedback Optimal Control \\ via Transformers Without Explicit State Estimation}

\maketitle

\begin{abstract}
Classical optimal control is used to design a separate controller for each plant. This paper asks a different question. Can one learned policy serve a whole family of partially observed, noisy linear systems? We show that one transformer can. The policy is trained to imitate optimal LQG control on a heterogeneous family of Linear Time-Invariant (LTI) systems with Multiple Inputs and Multiple Outputs (MIMO), whose dimensions, cost matrices, and noise levels all vary. A shared representation with system-wise standardization, zero-padding, masking, and an encoding of the cost and noise covariances lets one set of weights control every member of the family. At run time, a short window of noisy measurements is all the policy needs to produce a control action, with no plant matrices and no state estimator. We characterize the relation between the optimal LQG policy (represented as a dynamic state-space model) and a static transformer acting on a finite window of measurements. We then express the closed-loop excess cost of any causal policy as the energy of its deviation from the optimal action. Finally, we evaluate the approach on 28 simulated systems. The trained policy reaches the optimal cost to within a median gap of $0.4\%$ for seen systems. For unseen systems, a light fine-tuning step brings the median gap to $1.0\%$.
\end{abstract}

\section{Introduction}
\label{sec:introduction}
Linear Quadratic Gaussian (LQG) control is a standard optimal control method~\cite{anderson1990optimal}. This formulation is used to design output feedback controllers in the presence of sensor noise and process disturbance. The optimal controller is designed based on a (nominal) model of the plant dynamics. The optimal controller combines a Kalman filter~\cite{kalman1960new} with the Linear Quadratic Regulator (LQR) gain. A key issue is that the plant dynamics often deviate from the nominal design model due to operating-point shifts, environmental effects, and system aging / wear. Robust control~\cite{zhou1996robust} was developed to handle such model errors, but these methods can be conservative if the model variation is large. Alternatively, a controller can be designed for each plant. For example, each possible plant is identified via experimental data. An LQG controller is then designed for the identified plant model. We ask whether a single learned policy can replace this per-plant design. Specifically, can one transformer map a short window of noisy output measurements to a near-optimal control action across an entire family of linear systems?

Our earlier work answered this question in the deterministic full-state setting. In~\cite{binmohaya2026transformers}, a single transformer imitates optimal LQR state feedback for a set of MIMO LTI systems whose state dimensions, input dimensions, and cost objectives differ. That policy reads a window of past states and the cost matrices, and it needs no plant matrices at run time. A companion letter~\cite{binmohaya2026certification} supplies closed-loop certificates for that policy. Both works assume a noise-free plant whose state is measured exactly. This paper removes both assumptions.

Output feedback changes the learning problem in three ways. First, the optimal controller is no longer static. The LQR action depends only on the current state, whereas the LQG action depends on the entire measurement history through the filter. A policy with a finite window must therefore perform implicit filtering in addition to implicit system identification. Second, the windowed measurements are noisy. The dynamics must be inferred from data corrupted by the same disturbances that the controller is rejecting. Third, the optimal action depends on the noise statistics as well as on the cost. The same plant and cost call for a different controller when the sensors are less reliable.

\begin{figure}[t]
    \centering
    \includegraphics[width=\linewidth]{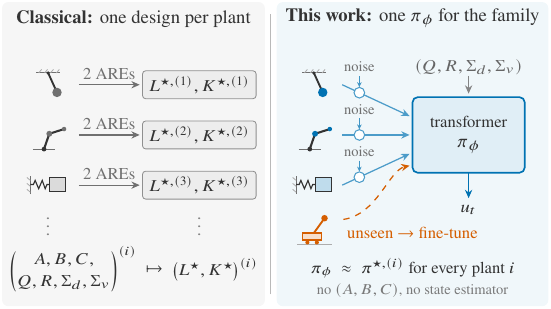}
    \caption{One transformer policy $\pi_\phi$ replaces the per-plant LQG design of a filter gain $L^\star$ and a control gain $K^\star$. It maps a window of noisy outputs and the matrices $(Q,R,\Sigma_d,\Sigma_v)$ to the control $u_t$.}
    \label{fig:overview}
            \vspace{-10pt}
\end{figure}

We cast cross-system LQG control as sequence-to-control learning as shown in Figure~\ref{fig:overview}. A single transformer policy is trained on closed-loop rollouts of optimal LQG controllers over many systems, cost settings, and noise levels. At each sample, the policy computes the control action from the window of recent measurements together with encodings of the cost matrices and the noise covariances. It uses no plant matrices, no explicit state estimator, and no Riccati solution at run time. Sequence models are a natural fit for this task. Recurrent architectures~\cite{hochreiter1997long} carry memory in a hidden state but are difficult to train through time~\cite{pascanu2013difficulty}. Transformers~\cite{vaswani2017attention} let all elements of the measurement window interact directly through self-attention. They train in parallel on millions of short windows from many systems.

\subsection{Related Work}
\label{sec:related_work}
The LQG controller is optimal for a known model, but it comes with no guaranteed robustness margins~\cite{doyle1978guaranteed}. Robust control~\cite{zhou1996robust} addresses model uncertainty, but the designs can be conservative if there is a large uncertainty set. Gain-scheduled design~\cite{shamma1988gain} addresses variations in plant dynamics across an operating envelope, but in principle this requires separate designs for every plant (or grid of plants). In all of these model-based methods, the design is repeated whenever the model, the cost, or the noise statistics change.

A learning-based line of work treats the unknown partially observed plant directly. Certainty-equivalent design identifies a model and then solves the Riccati equations, with a sub-optimality that scales with the square of the identification error~\cite{mania2019certainty}. Sample-complexity~\cite{zheng2021sample} and regret guarantees~\cite{lale2020logarithmic,simchowitz2020improper} are available for a single unknown system. A second route skips the model and optimizes the controller parameters directly. Policy optimization for LQG has a more intricate landscape than its LQR counterpart~\cite{fazel2018global,tang2023analysis}. Policy gradient can still be applied after the dynamic controller is rewritten as a static gain acting on a finite input-output history~\cite{sadamoto2025policy}. A similar lifting extends it to a collection of LQG tasks~\cite{toso2026multitask}. In all of these, the learned object is a model or a controller for one plant, or a shared controller whose sub-optimality grows with task heterogeneity. The state, input, and output dimensions are fixed throughout.

Generalist neural policies provide an alternative to LQG designs. Such policies pursue breadth across platforms. Policies trained with dynamics randomization use memory or an explicit adaptation module to infer unobserved plant parameters from recent observations~\cite{peng2018sim,kumar2021rma}. Transformers trained across embodiments~\cite{brohan2023rt1,Doshi24-crossformer}, and foundation models for low-level dynamics and control~\cite{xiao2024anycaranywherelearninguniversal,eschmann2025raptor}, carry this idea to many robots. These policies are trained by reinforcement learning or by imitation of human or learned teachers. None of them distills optimal output-feedback control into one policy or measures performance against a known optimum. Closer to this paper, transformers have been studied as estimators and controllers for linear systems. One transformer trained across random LTI systems learns output prediction in context and approaches the performance of the Kalman filter~\cite{du2023can,akram2024transformers}. In-context learning has also produced class models for system identification~\cite{forgione2023from}. On the representation side, an explicit transformer construction approximates the Kalman filter, and its use in measurement feedback approximates the LQG controller~\cite{goel2024can}. That result assumes a known system and establishes what a transformer can represent. Transformers have also been used to cast decision making as sequence modeling~\cite{chen2021decision} and to learn decision rules in context by supervised pretraining~\cite{lee2023supervised}.

Two gaps remain. First, no existing method distills optimal output-feedback control for a family of heterogeneous systems into one policy and evaluates it in closed loop against each instance's own LQG optimum. Second, no analysis relates a static policy on a finite measurement window to the dynamic LQG controller and to the closed-loop excess cost. This paper addresses both gaps.

\subsection{Contributions and Organization}
The paper has three main contributions. First, we extend the results in~\cite{binmohaya2026transformers} to output feedback. The expert and the learned policy read the same measurements before each action, so the LQG optimum is an appropriate reference for the learned closed loop. The policy reads standardized, zero-padded, and masked measurements, conditioned on the cost matrices and the noise covariances. The problem is formulated in \autoref{sec:problem_formulation} and the proposed method is described in \autoref{sec:method}. Second, we characterize the class of finite-window controllers (\autoref{subsec:window}). We show that the optimal LQG action is a linear function of the measurement window plus a remainder governed by the dynamics of the optimal controller. Third, we analyze the closed-loop excess cost (\autoref{subsec:excess}). We use a classical completion-of-squares identity to write the excess cost of any causal output-feedback policy as the weighted energy of its deviation from the optimal action along its own closed loop. The identity is classical. We use it to link the imitation error to the closed-loop cost and to obtain a per-rollout measure of sub-optimality that is nonnegative by construction. In addition, we validate the approach on 28 simulated systems (\autoref{sec:experiments}). For \textcolor{myblue}{seen systems}, the median gap to the optimal cost is $0.4\%$. For \textcolor{orange}{unseen systems} it is $1.0\%$ after fine-tuning.

\section{Problem Formulation}
\label{sec:problem_formulation}

Consider the discrete-time LTI system
\dsmall \begin{align}
    x_{t+1} &= A x_t + B u_t + d_t,
    \label{eq:system_state}\\
    y_t &= C x_t + v_t,
    \label{eq:system_output}
\end{align} \normalsize
where $x_t \in \R^{n_x}$ is the state, $u_t \in \R^{n_u}$ is the control input, and $y_t \in \R^{n_y}$ is the measured output at time $t$. The process disturbance $d_t \in \R^{n_x}$ and the measurement noise $v_t \in \R^{n_y}$ are independent and identically distributed processes. They are zero-mean Gaussian with covariances
\dsmall \begin{align}
    \mathbb{E}\bigl[d_t d_t^\top\bigr] = \Sigma_d \succeq 0,
    \qquad
    \mathbb{E}\bigl[v_t v_t^\top\bigr] = \Sigma_v \succ 0.
\end{align} \normalsize
They are independent of each other and of the initial state $x_0\sim\mathcal{N}(0,\Sigma_0)$. This is the standard LQG setting~\cite{anderson1990optimal}. The state is not measured. The control $u_t$ is computed before the output $y_t$ is processed. At time $t$ the controller therefore knows the past outputs $y_0,\ldots,y_{t-1}$ and its own past inputs $u_0,\ldots,u_{t-1}$. A causal output-feedback policy $\pi=\{\pi_t\}_{t\ge0}$ maps this information to the control, $u_t=\pi_t(y_0,\ldots,y_{t-1})$. The past inputs are functions of the past outputs under $\pi$, so they are not listed as separate arguments. The policy has to stabilize the plant while minimizing the infinite-horizon average cost
\dsmall \begin{align}
    \mathcal{J}(\pi)
    :=
    \lim_{T\to\infty}
    \frac{1}{T}\,
    \mathbb{E}
    \left[
    \sum_{t=0}^{T-1}
    x_t^\top Q x_t + u_t^\top R u_t
    \right],
    \label{eq:cost}
\end{align} \normalsize
where $Q \succeq 0$ and $R \succ 0$, and the expectation is over the disturbance, the noise, and the initial state. Unlike the deterministic LQR cost, $\mathcal{J}(\pi)$ does not depend on the initial condition, whose contribution vanishes in the time average.

We assume that (i) $(A,B)$ is stabilizable, (ii) $(Q^{1/2},A)$ is detectable, (iii) $(C,A)$ is detectable, and (iv) $(A,\Sigma_d^{1/2})$ is stabilizable. This is the infinite-horizon, discrete-time LQG problem~\cite{anderson1990optimal}. The assumptions ensure that the optimal controller exists and stabilizes the plant.

By the separation principle, the optimal policy $\pi^\star$ applies the LQR gain to a state estimate $\hat{x}_t$,
\dsmall \begin{align}
    u_t^\star = -K^\star \hat{x}_{t},
    \qquad
    K^\star = \bigl(R + B^\top P B\bigr)^{-1} B^\top P A,
    \label{eq:lqg_control}
\end{align} \normalsize
where $P$ solves the control Algebraic Riccati Equation (ARE)
\dsmall \begin{align}
    P = Q + A^\top P A
    - A^\top P B \bigl(R + B^\top P B\bigr)^{-1} B^\top P A.
    \label{eq:control_are}
\end{align} \normalsize
The estimate $\hat{x}_t$ uses the outputs $y_0,\ldots,y_{t-1}$. It is generated by the steady-state Kalman filter
\dsmall \begin{align}
    \hat{x}_{t+1}
    = A \hat{x}_{t} + B u_t
    + L^\star \bigl(y_{t} - C \hat{x}_{t}\bigr),
    \qquad \hat{x}_{0} = 0,
    \label{eq:observer}
\end{align} \normalsize
with the filter gain
\dsmall \begin{align}
    L^\star = A \Sigma C^\top \bigl(C \Sigma C^\top + \Sigma_v\bigr)^{-1},
    \label{eq:kalman_gain}
\end{align} \normalsize
where $\Sigma$ is the steady-state covariance of the estimation error $x_t-\hat{x}_t$. It solves the dual ARE
\dsmall \begin{align}
    \Sigma
    =
    A \Sigma A^\top + \Sigma_d
    - A \Sigma C^\top \bigl(C \Sigma C^\top + \Sigma_v\bigr)^{-1} C \Sigma A^\top.
    \label{eq:filter_are}
\end{align} \normalsize
The optimal cost is (see, e.g.,~\cite{anderson1990optimal})
\dsmall \begin{align}
    \mathcal{J}(\pi^\star)
    = \mathrm{tr}\bigl(P\Sigma_d\bigr)
    + \mathrm{tr}\bigl(\mathcal{W}K^\star\Sigma K^{\star\top}\bigr),
    \label{eq:optimal_cost}
\end{align} \normalsize
with $\mathcal{W}:=R+B^\top PB$. The first term is the cost of optimal state feedback under the disturbance, and the second is the price of partial observation.

The expert in \eqref{eq:lqg_control} acts on $y_0,\ldots,y_{t-1}$, so $u_t^\star=\pi^\star_t(y_0,\ldots,y_{t-1})$. The learned policy defined below also uses the plant measurements, but only over a finite window. The closed loop with the optimal LQG policy, which uses all past measurement history, serves as a reference for the learned closed loop. The impact of using a finite measurement window in the learned policy is discussed in \autoref{subsec:window}.

Consider $N$ such LQG problems,
\dsmall \begin{align}
    \left\{
    A^{(i)}, B^{(i)}, C^{(i)}, Q^{(i)}, R^{(i)}, \Sigma_d^{(i)}, \Sigma_v^{(i)}, \pi^{\star,(i)}
    \right\}_{i=1}^N,
\end{align} \normalsize
whose state, input, and output dimensions $(n_x^{(i)},n_u^{(i)},n_y^{(i)})$ may differ, and where $\pi^{\star,(i)}$ is the optimal controller of problem $i$. The superscript identifies the problem instance. We omit it whenever one instance is fixed.

We seek one universal policy
\dsmall \begin{align}
    u_t = \pi_\phi\left(\{y_l\}_{l=t-w-1}^{t-1},\,Q,\,R,\,\Sigma_d,\,\Sigma_v\right),
    \label{eq:policy}
\end{align} \normalsize
with parameter vector $\phi \in \R^{n_\phi}$. It must stabilize every instance and come close to the optimal cost of each. The policy is a static map of the $w+1$ most recent past outputs, where $w\in\mathbb{N}$ sets the window length. It also receives the cost matrices and the noise covariances of the instance. Both are known to the designer. The policy receives no plant matrices $(A,B,C)$, no state estimate, and no past controls.

The design goal is the smallest total closed-loop sub-optimality over the $N$ instances,
\dsmall \begin{align}
    \argmin_{\phi}\
    \sum_{i=1}^N
    \left[
    \mathcal{J}^{(i)}\bigl(\pi_\phi\bigr)
    -
    \mathcal{J}^{(i)}\bigl(\pi^{\star,(i)}\bigr)
    \right],
    \label{eq:primary_objective}
\end{align} \normalsize
where $\mathcal{J}^{(i)}$ is the LQG cost \eqref{eq:cost} of problem $i$. Every term is nonnegative, because $\pi^{\star,(i)}$ is optimal among all causal output-feedback policies and $\pi_\phi$ is one of them. We seek a policy $\pi_\phi$ that makes \eqref{eq:primary_objective} as small as possible. This policy also initializes the fine-tuned policies for \textcolor{orange}{unseen systems}.

\section{Method}
\label{sec:method}
We learn the universal policy \eqref{eq:policy} by supervised imitation of optimal LQG controllers. Closed-loop rollouts of these controllers provide the training data. A shared representation brings the data of all instances to one standardized input format. A transformer realizes the policy, and \textcolor{orange}{unseen systems} are handled by fine-tuning. A fuller account of the representation, the architecture, and the training is given in~\cite{binmohaya2026transformers}.

Fix an LQG problem $i\in\{1,\ldots,N\}$ with data $(A^{(i)}, B^{(i)}, C^{(i)}, Q^{(i)}, R^{(i)}, \Sigma_d^{(i)}, \Sigma_v^{(i)})$. We simulate the plant \eqref{eq:system_state} and \eqref{eq:system_output} in closed loop with its optimal controller $\pi^{\star,(i)}$ for $M$ independent draws of the initial state, the disturbances, and the measurement noise. This yields the length-$T_{\mathrm{tr}}$ ($T_{\mathrm{tr}}\in\mathbb{N}$) output and control sequences $\{y_t^{(i,j)}\}_{t=0}^{T_{\mathrm{tr}}-1}$ and $\{u_t^{(i,j)}\}_{t=0}^{T_{\mathrm{tr}}-1}$, $j=1,\ldots,M$. The recorded control $u_t^{(i,j)}$ is a function of $y_0^{(i,j)},\ldots,y_{t-1}^{(i,j)}$. These outputs are available to the learned policy at time $t$. The states and the state estimates are used only inside the simulation and are not recorded. The $N$ instances are drawn from mechanical, electrical, and robotic LTI systems, so that the training family is broad. We do not vary the plant parameters. The process disturbance and the measurement noise already drive every rollout away from the nominal trajectories. The noise covariances are fixed per system. An instance is the full tuple of plant, cost, and noise data. Changing the cost alone therefore yields a new instance with a different optimal controller. The cost changes the gain $K^\star$, and the noise covariances change the filter gain $L^\star$. We therefore vary the cost settings of every system. The policy receives $(Q,R,\Sigma_d,\Sigma_v)$ as input. It therefore learns how the optimal feedback depends on the cost and noise specification, instead of memorizing one fixed design.

The data of different instances live in different spaces and on different scales. We bring them to one input format in four steps. In the first step, every system is standardized using its own statistics. We fix one problem $i$ and omit its superscript. Let $\mu_y,\sigma_y\in\R^{n_y}$ be the entrywise mean and standard deviation of the recorded outputs, and let $\mu_u,\sigma_u\in\R^{n_u}$ be the same statistics of the recorded controls. The statistics are pooled over all training trajectories of that system and are computed once. The standardized output and control of trajectory $j$ at time $t$ are
\dsmall \begin{align}
\tilde{y}_t^{(j)} &= (y_t^{(j)}-\mu_y)\oslash\sigma_y, \\
\tilde{u}_t^{(j)} &= (u_t^{(j)}-\mu_u)\oslash\sigma_u,
\end{align} \normalsize
where $\oslash$ is entrywise division and every entry of $\sigma_y$ and $\sigma_u$ is positive. The same statistics are used at inference to map the predicted control back to physical units.

In the second step, all dimensions are brought to common sizes by zero-padding. Let $\mathrm{Pad}(\cdot,\,n)$ denote the operator that extends its argument with trailing zeros to dimension $n$, and let $n_x^{\mathrm{max}}:=\max_i n_x^{(i)}$, $n_u^{\mathrm{max}}:=\max_i n_u^{(i)}$, and $n_y^{\mathrm{max}}:=\max_i n_y^{(i)}$ be the largest dimensions in the family. The padded data are
\dsmall \begin{align}
\bar{y}_t^{(j)} = \mathrm{Pad}\big(\tilde{y}_t^{(j)}, n_y^{\mathrm{max}}\big),
\qquad
\bar{u}_t^{(j)} = \mathrm{Pad}\big(\tilde{u}_t^{(j)}, n_u^{\mathrm{max}}\big).
\end{align} \normalsize
For $t\le w$ the missing past outputs are also zero-padded, so that every window has length $w+1$.

In the third step, the objective and the noise level are encoded. In all instances considered here these matrices are diagonal with positive entries,
$Q=\mathrm{diag}(q)$, $R=\mathrm{diag}(r)$, $\Sigma_d=\mathrm{diag}(s_d)$, and $\Sigma_v=\mathrm{diag}(s_v)$, with $q,s_d\in\R^{n_x}_{>0}$, $r\in\R^{n_u}_{>0}$, and $s_v\in\R^{n_y}_{>0}$. The four vectors $(q,r,s_d,s_v)$ thus specify the objective and the noise statistics exactly. Their entries span many orders of magnitude across systems. We therefore apply the logarithm entrywise, which brings them to a comparable scale, and then zero-pad,
\dsmall \begin{align}
\bar{q} &:= \mathrm{Pad}\big(\log q,\, n_x^{\mathrm{max}}\big), &
\bar{r} &:= \mathrm{Pad}\big(\log r,\, n_u^{\mathrm{max}}\big), \nonumber\\
\bar{s}_d &:= \mathrm{Pad}\big(\log s_d,\, n_x^{\mathrm{max}}\big), &
\bar{s}_v &:= \mathrm{Pad}\big(\log s_v,\, n_y^{\mathrm{max}}\big).
\end{align} \normalsize
The encodings form the vector $c:=\bar{q}\oplus\bar{r}\oplus\bar{s}_d\oplus\bar{s}_v$ of dimension $d_c=2n_x^{\mathrm{max}}+n_u^{\mathrm{max}}+n_y^{\mathrm{max}}$, where $\oplus$ denotes vector concatenation. The vector $c$ is attached to every padded output in the window, which gives the input
\dsmall \begin{align}
S_t^{(j)} = \begin{bmatrix} (\bar{y}_{t-w-1}^{(j)})^\top \oplus c^\top \\ \vdots\\ (\bar{y}_{t-1}^{(j)})^\top \oplus c^\top \end{bmatrix} \label{eq:S}
\end{align} \normalsize
of fixed size $(w+1)\times d_{\mathrm{in}}$ with $d_{\mathrm{in}}=n_y^{\mathrm{max}}+d_c$. The entries $\bar{q}$ and $\bar{s}_d$ refer to state coordinates that the policy never observes. They act as a descriptor of the objective and of the disturbance level within a system family, and the policy learns their effect on the optimal action from data.

In the fourth step, padded control entries are excluded from the loss by the binary mask $\kappa := \mathrm{Pad}(1_{n_u},\, n_u^{\mathrm{max}})$, where $1_n\in\R^n$ is the all-ones vector. A training sample is the triplet
$(S_t^{(i,j)},\bar{u}_t^{(i,j)},\kappa^{(i)})$,
with the current control as the target. The dataset $\mathcal{D}$ collects these triplets over all $t\in\{0,\dots,T_{\mathrm{tr}}-1\}$, trajectories $j\in\{1,\dots,M\}$, and problems $i\in\{1,\dots,N\}$.

On the standardized data, the transformer implements the map $\bar{\pi}_\phi:\R^{(w+1)\times d_{\mathrm{in}}}\to\R^{n_u^{\mathrm{max}}}$ with parameter vector $\phi$. Its input is $S_t$ from \eqref{eq:S}, with the superscript $j$ omitted, and its output is the predicted standardized control $\hat{\bar{u}}_t=\bar{\pi}_\phi(S_t)$. We use the architecture of~\cite{binmohaya2026transformers} and summarize it here. The rows of $S_t$ are embedded into width $d_{\mathrm{m}}$ by an affine map, and a learned positional matrix $\mathcal{P}\in\R^{(w+1)\times d_{\mathrm{m}}}$ is added to encode temporal order,
\dsmall \begin{align}
H = S_tW_{\mathrm{in}}^{\top} + 1_{w+1}b_{\mathrm{in}}^{\top} + \mathcal{P}\ \in\ \R^{(w+1)\times d_{\mathrm{m}}},
\label{eq:H}
\end{align} \normalsize
where $W_{\mathrm{in}}\in\R^{d_{\mathrm{m}}\times d_{\mathrm{in}}}$ and $b_{\mathrm{in}}\in\R^{d_{\mathrm{m}}}$. A transformer block maps $H$ to an output of the same size. It applies multi-head self-attention~\cite{vaswani2017attention} with $h$ heads, a residual connection followed by layer normalization, a position-wise feed-forward network of width $d_{\mathrm{ff}}$ with GELU activation~\cite{hendrycks2016gaussian}, and a second residual connection followed by layer normalization~\cite{vaswani2017attention}. We denote this map by $\mathrm{Block}(H)$. The policy uses $n_{\mathrm{b}}\in\mathbb{N}$ such blocks, each with its own parameters, applied in parallel to the same embedded sequence,
\dsmall \begin{align}
Z^{(\ell)} = \mathrm{Block}_{\ell}\big(H\big) \in \R^{(w+1)\times d_{\mathrm{m}}},
\quad \ell=1,\ldots,n_{\mathrm{b}}.
\end{align} \normalsize
The last row of each block output corresponds to the current time step. We write it as a column vector $z^{(\ell)}\in\R^{d_{\mathrm{m}}}$. These vectors are stacked and passed through one linear readout that produces the control,
\dsmall \begin{align}
\hat{\bar{u}}_t = \bar{\pi}_\phi(S_t) = W_{\mathrm{out}}\big[z^{(1)};z^{(2)};\dots;z^{(n_{\mathrm{b}})}\big] + b_{\mathrm{out}} \in \R^{n_u^{\mathrm{max}}},
\end{align} \normalsize
with $W_{\mathrm{out}}\in\R^{n_u^{\mathrm{max}}\times (n_{\mathrm{b}}d_{\mathrm{m}})}$ and $b_{\mathrm{out}}\in\R^{n_u^{\mathrm{max}}}$. All weights of the embedding, the positional matrix, the blocks, and the readout together form the parameter vector $\phi$. Self-attention lets the representation of the current time step weight the outputs in the window in a data-dependent way. This gives a single set of weights the flexibility to realize instance-dependent linear maps from the window to the action.

The parameters are fitted by minimizing a masked Cauchy loss over mini-batches $\mathcal{B}$ of sample indices $(i,j,t)$ drawn from $\mathcal{D}$,
\dsmall \begin{align}
\mathcal{L}
= \frac{1}{|\mathcal{B}|}\!\sum_{(i,j,t) \in \mathcal{B}}\!\!
\ln\!\Big(1+\tfrac{1}{\xi^2}\bigl\|\kappa^{(i)}\odot\bigl(\bar{\pi}_\phi(S_t^{(i,j)})-\bar{u}_t^{(i,j)}\bigr)\bigr\|_2^2\Big),
\label{eq:loss}
\end{align} \normalsize
where $\xi>0$ is a scale parameter and $\odot$ denotes the entrywise product. The mask removes the padded entries from the loss, and the logarithm limits the influence of large residuals. The optimizer is AdamW~\cite{loshchilov2018decoupled}. The design objective \eqref{eq:primary_objective} is not optimized directly. The supervised loss serves as its stand-in. \autoref{sec:analysis} relates the two.

After training, the policy is placed in closed loop with the plant. At each time step the window of past outputs is standardized, padded, and joined with the encoding $c$ to form $S_t$. The control is applied. The newest output $y_t$ is then appended to the window. The physical control for problem $i$ is recovered by de-standardizing the first $n_u^{(i)}$ entries of the prediction,
\dsmall \begin{align}
u_t^{(i)} = \sigma_u^{(i)}\odot\hat{\bar{u}}_t^{(i)}(1:n_u^{(i)}) + \mu_u^{(i)}.
\end{align} \normalsize
No state estimator runs alongside the policy, and no plant matrices are used. The only system-specific quantities at run time are the standardization statistics and the encoding $c$. During closed-loop operation, the window holds outputs generated by the learned policy rather than by the expert.

A system outside the training family is handled by fine-tuning. We collect a small set of expert rollouts on that system, compute its standardization statistics from these rollouts, and update the trained weights on the resulting samples with the same loss. The result is a copy of the policy that is specialized to the new system.

\section{Analysis}
\label{sec:analysis}

\subsection{What a Window of Outputs Can Represent}
\label{subsec:window}
The optimal LQG controller is a dynamic system, whereas the policy \eqref{eq:policy} sees only $w+1$ outputs. We now make precise which part of the optimal action is available in the window. Fix an instance. Along an expert rollout, substituting $u_{t-1}^\star=-K^\star\hat{x}_{t-1}$ into \eqref{eq:observer} gives a recursion driven by the outputs alone,
\dsmall \begin{align}
    \hat{x}_{t} = F\hat{x}_{t-1} + L^\star y_{t-1},
    \quad
    F := A-BK^\star-L^\star C.
    \label{eq:controller_recursion}
\end{align} \normalsize
Unrolling \eqref{eq:controller_recursion} over the window and applying \eqref{eq:lqg_control} yields
\dsmall \begin{align}
    u_t^\star
    = -K^\star\sum_{k=0}^{w} F^{k}L^\star\,y_{t-1-k}
    \;-\; K^\star F^{\,w+1}\,\hat{x}_{t-w-1}.
    \label{eq:unroll}
\end{align} \normalsize
The first term is a linear function of the window. Its coefficients $K^\star F^kL^\star$ depend on the instance, and the plant matrices that determine them are not given to the policy. The transformer must infer them from the window and from the encoding $c$ within the structured family it is trained on. This is the implicit identification step, and it now takes place on noisy data. The second term is the part of the optimal action that the window cannot see. For $t\le w+1$ there is no remainder. The filter starts from $\hat{x}_0=0$, so the sum stops at $k=t-1$. For $t>w+1$ the remainder decays geometrically in $w$ whenever the spectral radius of $F$ is below one.

The eigenvalues of $F$ are the poles of the optimal controller itself. They differ from the closed-loop poles, which are the eigenvalues of $A-BK^\star$ and of $A-L^\star C$. The window length $w$ trades approximation accuracy for computation, and the required length depends on the slowest controller mode. We treat $w$ as a design parameter.

Relation \eqref{eq:unroll} also clarifies the role of past controls. A window of past outputs and past controls is an alternative input. The policy would remain a static map on that window, but its own past actions would then enter its input. Its remainder is governed by $A-L^\star C$, which is stable under the standing assumptions, whereas the remainder in \eqref{eq:unroll} is governed by $F$. Along expert rollouts, past controls are determined by past outputs, so they add no information. We use outputs only, so that the policy never feeds back its own actions, and leave the comparison for future work.

\subsection{Closed-Loop Excess Cost}
\label{subsec:excess}
The training loss \eqref{eq:loss} measures the imitation error on expert rollouts, whereas the design objective \eqref{eq:primary_objective} concerns the cost of the learned closed loop. The following classical identity links the two. Fix an instance and write $\|z\|_{\mathcal{W}}^2:=z^\top\mathcal{W}z$, with $\mathcal{W}$ as in \eqref{eq:optimal_cost}.

\begin{proposition}[Exact excess cost]
\label{prop:excess}
Let $\pi$ be a causal output-feedback policy with $u_t=\pi_t(y_0,\ldots,y_{t-1})$. Assume that its closed loop is mean-square bounded, $\sup_t\mathbb{E}\|x_t\|_2^2<\infty$. Let $\hat{x}_{t}$ be generated by the Kalman filter \eqref{eq:observer} driven by the closed-loop outputs and the applied controls. Then
\dsmall \begin{align}
    \mathcal{J}(\pi)-\mathcal{J}(\pi^\star)
    = \lim_{T\to\infty}\frac{1}{T}\,
    \mathbb{E}\left[\sum_{t=0}^{T-1}\bigl\|u_t+K^\star\hat{x}_{t}\bigr\|_{\mathcal{W}}^2\right],
    \label{eq:excess}
\end{align} \normalsize
and the limit defining $\mathcal{J}(\pi)$ exists if and only if the limit in \eqref{eq:excess} exists.
\end{proposition}

\begin{proof}
The ARE \eqref{eq:control_are} gives, for all $x$ and $u$, the completion of squares
\dsmall \begin{align}
    x^\top Qx+u^\top Ru+(Ax+Bu)^\top P(Ax+Bu)-x^\top Px \nonumber\\
    =\|u+K^\star x\|_{\mathcal{W}}^2.
    \label{eq:squares}
\end{align} \normalsize
Since $d_t$ is zero-mean and independent of $(x_t,u_t)$, we have $\mathbb{E}[x_{t+1}^\top Px_{t+1}]=\mathbb{E}[(Ax_t+Bu_t)^\top P(Ax_t+Bu_t)]+\mathrm{tr}(P\Sigma_d)$. Evaluating \eqref{eq:squares} along the closed loop and averaging expectations over $t=0,\ldots,T-1$ gives
\dsmall \begin{align}
    &\frac{1}{T}\mathbb{E}\Bigl[\sum_{t=0}^{T-1}x_t^\top Qx_t+u_t^\top Ru_t\Bigr] \nonumber\\
    &\quad= \mathrm{tr}(P\Sigma_d)
    +\frac{1}{T}\mathbb{E}\Bigl[\sum_{t=0}^{T-1}\|u_t+K^\star x_t\|_{\mathcal{W}}^2\Bigr] \nonumber\\
    &\qquad+\frac{1}{T}\mathbb{E}\bigl[x_0^\top Px_0-x_T^\top Px_T\bigr].
    \label{eq:telescoped}
\end{align} \normalsize
The last term vanishes as $T\to\infty$ by mean-square boundedness. Next, write $u_t+K^\star x_t=(u_t+K^\star\hat{x}_{t})+K^\star e_t$ with $e_t:=x_t-\hat{x}_{t}$. Consider first $\Sigma_0=\Sigma$, with $\hat{x}_0=0=\mathbb{E}[x_0]$. The control enters $x_t$ and $\hat{x}_t$ through the same term $Bu_{t-1}$, so $e_t$ does not depend on the policy. It is the estimation error of the uncontrolled system, which is zero-mean Gaussian with covariance $\Sigma$ and independent of the past measurements~\cite{astrom1970introduction}. Hence $\hat{x}_t$ is the conditional mean of $x_t$ given $y_0,\ldots,y_{t-1}$ under any causal policy, with $\mathbb{E}[e_t\mid y_0,\ldots,y_{t-1}]=0$ and $\mathrm{Cov}(e_t)=\Sigma$, which is all the argument uses. The term $u_t+K^\star\hat{x}_{t}$ is a function of these outputs. The cross term therefore has zero mean, and
\dsmall \begin{align}
    \mathbb{E}\|u_t+K^\star x_t\|_{\mathcal{W}}^2 =\mathbb{E}\|u_t+K^\star\hat{x}_{t}\|_{\mathcal{W}}^2+\mathrm{tr}\bigl(\mathcal{W}K^\star\Sigma K^{\star\top}\bigr).
    \label{eq:orthogonality}
\end{align} \normalsize
Substituting \eqref{eq:orthogonality} into \eqref{eq:telescoped} and using \eqref{eq:optimal_cost} proves the claim. For $\Sigma_0\neq\Sigma$, the conditional mean comes from the time-varying Kalman filter, whose gain converges to $L^\star$. Their difference is independent of the policy and vanishes in mean square, so the time averages are unchanged.
\end{proof}

The windowed policy \eqref{eq:policy} is causal, so \autoref{prop:excess} applies to it whenever its closed loop is mean-square bounded. The proposition has two consequences for this paper. First, it identifies what imitation must achieve. The sub-optimality of a policy is exactly the $\mathcal{W}$-weighted energy of its deviation from the action that the optimal controller would take on the same closed-loop data. The training loss \eqref{eq:loss} penalizes this deviation in standardized coordinates and along the expert rollouts rather than the learned closed loop, so it is a surrogate and not a bound. Second, it provides a way to measure sub-optimality in simulation. The right-hand side of \eqref{eq:excess} averages squared norms, so the error energy sampled along any rollout is nonnegative and estimates the expected excess cost up to the terminal term in \eqref{eq:telescoped}. The difference between two noisy realized costs has no such property. Evaluating \eqref{eq:excess} requires the plant matrices in order to run the filter. These are available in simulation, and the policy itself never uses them.

        \vspace{-5pt}
\section{Experimental Validation}
\label{sec:experiments}
We evaluate the policy on the 28 simulated mechanical, electrical, and robotic LTI systems listed in \autoref{tab:lti_systems}. The first 15 systems are \textcolor{myblue}{seen} during training, and the remaining 13 are \textcolor{orange}{unseen}. The state dimension ranges from 2 to 12. The input and output dimensions range from 1 to 6. Each system measures only part of its state through $C$. The padded dimensions are $n_x^{\mathrm{max}}=12$, $n_u^{\mathrm{max}}=6$, and $n_y^{\mathrm{max}}=6$. All systems are discrete-time models with a sampling time of $0.02$ seconds. Initial states are drawn uniformly from a system-specific box. The covariance of this distribution is $\Sigma_0$. The noise covariances $\Sigma_d$ and $\Sigma_v$ are diagonal and fixed per system. Each system contributes nine cost settings from a $3\times3$ palette over $(Q,R)$. The first $Q$ and $R$ palettes are the nominal diagonals. The other palettes scale each diagonal entry by a factor drawn log-uniformly between $1/3$ and $3$.

\begin{table}[t]
\small
\centering
\caption{Benchmark systems with $(n_x,n_u,n_y)$.}
\label{tab:lti_systems}
\begin{tabular}{@{}>{\raggedright\arraybackslash}p{0.98\linewidth}@{}}
\toprule
\textbf{\textcolor{myblue}{Seen Systems:}}
\textbf{1.}~Simple Pendulum $(2,1,1)$, \textbf{2.}~Two Link Arm $(4,2,2)$, \textbf{3.}~Spring Damper $(2,1,1)$,
\textbf{4.}~Suspension System $(4,1,2)$, \textbf{5.}~DC Motor $(3,1,2)$, \textbf{6.}~Three Link Manipulator $(6,3,3)$,
\textbf{7.}~Differential Drive Robot $(5,2,3)$,
\textbf{8.}~SCARA Robot $(8,4,4)$, \textbf{9.}~Omnidirectional Robot $(6,3,3)$,
\textbf{10.}~Cable Driven Robot $(4,3,2)$, \textbf{11.}~Flexible Joint Robot $(4,1,2)$, \textbf{12.}~Six DOF Manipulator $(12,6,6)$,
\textbf{13.}~Dual Arm Robot $(8,4,4)$, \textbf{14.}~Double Integrator $(2,1,1)$, \textbf{15.}~Lotka Volterra $(2,2,2)$.\\
\midrule
\textbf{\textcolor{orange}{Unseen Systems:}}
\textbf{16.}~Inverted Pendulum $(2,1,1)$, \textbf{17.}~Segway Robot $(4,1,4)$, \textbf{18.}~Asymmetric Oscillator $(4,1,2)$,
\textbf{19.}~Active Mass Damper $(4,1,2)$, \textbf{20.}~Coupled Oscillators $(4,1,2)$, \textbf{21.}~Damped Oscillator $(2,1,1)$,
\textbf{22.}~Triple Mass Spring $(6,1,3)$, \textbf{23.}~Electromechanical Actuator $(3,1,2)$, \textbf{24.}~Thermal System $(3,1,3)$,
\textbf{25.}~Fluid Tank $(2,1,2)$, \textbf{26.}~Vibrating Beam $(4,1,4)$, \textbf{27.}~Motor Generator $(3,1,3)$,
\textbf{28.}~Mechanical Linkage $(4,1,2)$.\\
\bottomrule
\end{tabular}
        \vspace{-15pt}
\end{table}

Expert rollouts of the optimal controllers on the 15 \textcolor{myblue}{seen systems} last up to $T_{\mathrm{tr}}=1250$ steps and stop at settling. Rollouts are added until each system has 750 thousand training windows, 11.25 million in total. The window length is $w=19$ steps. Each window holds the 20 most recent past outputs. A trajectory-level split holds out $5\%$ of the data for validation, and windows from a validation trajectory never appear in training. The policy has four parallel transformer blocks with embedding width 64, eight attention heads, and feed-forward width 256, for about 204 thousand parameters. The Cauchy scale is $\xi=1$. Training uses AdamW with peak learning rate $5\times10^{-6}$, weight decay $10^{-5}$, gradient clipping at norm 2, and batch size 4096. It runs for two iterations of 500 epochs, about 12 hours on three NVIDIA A40 GPUs. For each \textcolor{orange}{unseen system} we fine-tune the model of the trained policy. The demonstrations are expert rollouts from its nine cost settings with six initial conditions each and 500 steps per rollout. Fine-tuning runs two epochs of Adam with learning rate $10^{-4}$ and batch size 256. The standardization statistics of an \textcolor{orange}{unseen system} come from the same rollouts. \textcolor{myblue}{Seen systems} use the shared policy and \textcolor{orange}{unseen systems} use their fine-tuned models.

Consider instance $i$ and one rollout of horizon $T_i$ under the learned policy. Let $\hat{x}_t$ be the estimate of the Kalman filter \eqref{eq:observer} driven by the outputs and the controls of this rollout. The \emph{relative sub-optimality} of the rollout is
\dsmall \begin{align}
\Delta^{(i)} := \frac{\sum_{t=0}^{T_i-1}\bigl\|u_t+K^\star\hat{x}_t\bigr\|_{\mathcal{W}}^2}{\sum_{t=0}^{T_i-1} x_t^{\star\top} Q x_t^\star + u_t^{\star\top} R u_t^\star},
\label{eq:metric}
\end{align} \normalsize
where $(x_t^\star,u_t^\star)$ is the rollout of the optimal controller from the same initial state and with the same disturbance and noise realizations. The numerator is the sampled right-hand side of \eqref{eq:excess}, so $\Delta^{(i)}$ is nonnegative by construction. The denominator is the realized optimal cost over the same horizon. Each system is evaluated on its nine trained cost settings with 25 random initial conditions each. This gives 225 rollouts per system and 6300 rollouts in total. Each rollout runs for $T_i$ steps, at which the optimal rollout settles, or $T_{\max}=500$ when it does not settle within that time. The optimal rollout settles when its state changes by less than $5\times10^{-3}$ per step for 15 consecutive steps.

\begin{figure}[t]
    \centering
    \includegraphics[width=\linewidth]{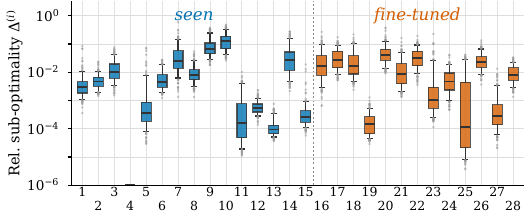}
    \caption{Relative sub-optimality $\Delta^{(i)}$ of \eqref{eq:metric} for each system in \autoref{tab:lti_systems}. Boxes span the interquartile range. The line inside each box is the median. Whiskers extend from the 5th to the 95th percentile. Dots are outliers. Blue boxes show the \textcolor{myblue}{seen systems} and orange boxes show the \textcolor{orange}{unseen systems} after fine-tuning. The values of the suspension system (\textcolor{myblue}{seen $\#4$}) lie below $10^{-6}$, so its box collapses onto the lower limit of the axis.}
    \label{fig:relative}
        \vspace{-15pt}

\end{figure}

The distribution of $\Delta^{(i)}$ over the rollouts of every system is shown in \autoref{fig:relative}. The Segway (\textcolor{orange}{unseen $\#17$}) and the vibrating beam (\textcolor{orange}{unseen $\#26$}) are the most challenging systems. They required a small extra fine-tuning budget, about 7 minutes each on an Apple M4 chip, against 5 minutes for all others. Their medians are $2.7\%$ and $2.3\%$. On the \textcolor{myblue}{seen systems} the pooled median is $0.4\%$. On the \textcolor{orange}{unseen systems} the pooled median after fine-tuning is $1.0\%$. Over all 6300 rollouts of the 28 systems the pooled median is $0.6\%$. No divergence is observed within the evaluation in any of the rollouts.

\begin{figure}[t]
    \centering
    \includegraphics[width=0.85\linewidth]{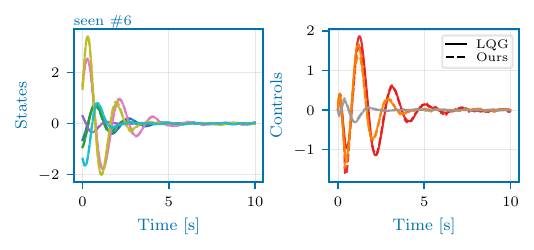}\\[2pt]
    \includegraphics[width=0.85\linewidth]{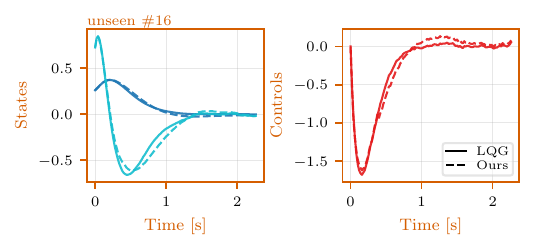}
    \caption{Closed-loop rollouts on the three-link manipulator (\textcolor{myblue}{seen $\#6$}, top) and the inverted pendulum (\textcolor{orange}{unseen $\#16$}, fine-tuned, bottom). Solid lines show the optimal LQG controller. Dashed lines show the learned policy. Both receive identical noise.}
    \label{fig:traj}
    \vspace{-10pt}
\end{figure}

\autoref{fig:traj} shows representative closed-loop rollouts. Each panel compares the learned policy with the optimal LQG controller. Both controllers start from the same initial state and receive identical process and measurement noise. The left plot shows the states, the right plot the controls. In the top panel of \autoref{fig:traj}, the base (before fine-tuning) policy reproduces the optimal output-feedback behavior on the three-link manipulator (\textcolor{myblue}{seen $\#6$}) almost exactly. In the bottom panel, the fine-tuned policy stabilizes the inverted pendulum (\textcolor{orange}{unseen $\#16$}) and tracks the optimal trajectories closely.

\section{Conclusions}
\label{sec:conclusions}
This paper showed that a single transformer policy can serve a family of partially observed, noisy linear systems. The policy imitates optimal LQG controllers across systems with different dimensions, costs, and noise levels. At run time, the only inputs are a short window of noisy measurements and an encoding of $(Q,R,\Sigma_d,\Sigma_v)$. No plant matrices and no state estimator are needed. A finite-window analysis identified which part of the optimal action the window can represent. Moreover, we provided an exact identity to express the closed-loop excess cost as the energy of the deviation from the optimal action. Across 28 simulated systems, the median gap to the optimal cost is $0.4\%$ for \textcolor{myblue}{seen systems} and $1.0\%$ for \textcolor{orange}{unseen systems} after fine-tuning. Over all rollouts it is $0.6\%$. No divergence is observed within the evaluation horizon. Future work includes closed-loop stability and near-optimality theoretical certificates in this output-feedback setting.

\section*{Statement on AI usage}
Generative AI tools were used only for editing and proofreading. All technical content and conclusions were developed and verified by the authors.

\printbibliography
\end{document}